\documentclass[runningheads]{llncs}
\usepackage{graphicx}
\usepackage[utf8]{inputenc}
\usepackage{amsmath}
\usepackage{amssymb}
\usepackage{stmaryrd}

\usepackage{booktabs}

\usepackage{xcolor}
\definecolor{darkgreen}{rgb}{0.0, 0.6, 0.0}
\definecolor{teal700}{HTML}{00796B}
\usepackage[hidelinks, colorlinks, citecolor=darkgreen]{hyperref}
\usepackage{cleveref}

\usepackage{tikz}
\usetikzlibrary{decorations.pathmorphing,shapes.misc,positioning}

\usepackage{macros}

\usepackage{orcidlink}
\renewcommand\orcidID[1]{\orcidlink{#1}}

\begin{document}
\title{Peregrine 2.0: Explaining Correctness of Population Protocols through Stage Graphs%
\thanks{This project has received funding from the European Research Council (ERC) under the European Union's Horizon 2020 research and innovation programme under grant agreement No~787367 (PaVeS). We thank Michael Blondin for contributions to the frontend and Philip Offtermatt for improvements of the simulation backend.}}

\titlerunning{Peregrine 2.0: Explaining Correctness through Stage Graphs}

\author{Javier Esparza\orcidID{0000-0001-9862-4919} \and
Martin Helfrich\orcidID{0000-0002-3191-8098} \and
Stefan Jaax\orcidID{0000-0001-5789-8091} \and
Philipp J. Meyer\orcidID{0000-0003-1334-9079}}
\authorrunning{Esparza et al.}

%
\institute{Technical University of Munich, Germany \\
\email{\{esparza,helfrich,jaax,meyerphi\}@in.tum.de}}
\maketitle              
\begin{abstract}
We present a new version of Peregrine, the tool for the
analysis and parameterized verification of population protocols
introduced in [Blondin et al., CAV'2018].
Population protocols are a model of computation, intensely studied by the
distributed computing community, in which mobile anonymous agents
interact stochastically to perform a task.

Peregrine 2.0 features a novel verification engine based on
the construction of stage graphs. Stage graphs are proof
certificates, introduced in [Blondin et al., CAV'2020], that are
typically succinct and can be independently checked. Moreover, unlike
the techniques of Peregrine 1.0, the stage graph methodology
can verify protocols whose executions never terminate, a class
including recent fast majority protocols.
Peregrine 2.0 also features a novel proof visualization
component that allows the user to interactively explore the stage
graph generated for a given protocol.

\end{abstract}

\section{Introduction}
We present Peregrine 2.0\footnote{
Peregrine 2.0 is available at \url{https://peregrine.model.in.tum.de/}.},
a tool for analysis and parameterized verification of
population protocols. Population protocols are a model of computation, intensely
studied by the distributed computing community, in which an arbitrary number
of indistinguishable agents interact stochastically in order to decide a given property 
of their initial configuration. For example, agents could initially be in one
of two possible states, “yes” and “no”, and their task could consist of deciding
whether the initial configuration has a majority of “yes” agents or not.

Verifying correctness and/or efficiency of a protocol is a very hard problem,
because the semantics of a protocol is an infinite collection of finite-state Markov
chains, one for each possible initial configuration. Peregrine 1.0 \cite{peregrine} was the first
tool for the automatic verification of population protocols. It relies on theory
developed in \cite{BlondinEJM17}, and is implemented on top of the Z3 SMT-solver.

Peregrine 1.0 could only verify protocols whose agents eventually never change
their state (and not only their answer). This constraint has become increasingly
restrictive, because it is not satisfied by many efficient and succinct protocols 
recently developed for different tasks \cite{AlistarhG18,BlondinEJ18,BlondinEGHJ20}. 
Further, Peregrine 1.0 was unable to provide correctness certificates and the user had to trust the tool.
Finally, Peregrine 1.0 did not provide any support for computing parameterized
bounds on the expected number of interactions needed to reach a stable consensus,
 i.e., bounds like ``$\lO ( n^2 \log n )$ interactions, where $n$ is the number of agents''.

Peregrine 2.0 addresses these three issues. It features a novel verification
engine based on theory developed in \cite{BEK18,CAV20}, which, given a protocol and a task 
description, attempts to construct a stage graph. Stage graphs are proof certificates
that can be checked by independent means, and not only prove the protocol
correct, but also provide a bound on its expected time-to-consensus. Stages 
represent milestones reached by the protocol on the way to consensus. Stage graphs
are usually small, and help designers to understand why a protocol works. The
second main novel feature of Peregrine 2.0 is a visualization component that offers 
a graphical and explorable representation of the stage graph.

The paper is organized as follows. Section \ref{sec:prelims} introduces population protocols
and sketches the correctness proof of a running example. Section \ref{sec:verif} describes the stage graph
generated for the example by Peregrine 2.0, and shows that it closely matches the human proof. 
Section \ref{sec:visual} describes the visualization component.

\section{Population protocols}\label{sec:prelims}

A \emph{population protocol} consists of a set $Q$ of \emph{states} with a subset 
$I \subseteq Q$ of \emph{initial states}, a set $T \subseteq Q^2 \times Q^2$ of \emph{transitions},
and an \emph{output function} $O: Q \rightarrow \set{0,1}$ assigning to each state a boolean output.
Intuitively, a transition $q_1, q_2 \mapsto q_3, q_4$ means that two agents in states $q_1, q_2$ can interact and
simultaneously move to states $q_3, q_4$. A \emph{configuration} is a mapping $C: Q \rightarrow \N$, where $C(q)$
represents the number
of agents in a state $q$.
An \emph{initial configuration} is a mapping $C: I \rightarrow \N$. A configuration has \emph{consensus $b \in \{0,1\}$} if all agents are in states with
output $b$.
We write configurations using a set-like notation, e.g. $C = \multiset{\qy,\qn,\qn}$ or $C = \multiset{\qy,2\cdot \qn}$ is the configuration where $C(\qy) = 1$, $C(\qn) = 2$ and $C(q) = 0$ for $q \not\in \{\qy,\qn\}$.

\paragraph{Running example: Majority Voting.}
The goal of this protocol is to conduct a vote by majority in a distributed way.
The states are $\set{\qY,\qN,\qy,\qn}$.  Initially, all agents are in state $\qY$ or $\qN$, 
according to how they vote.  The goal of the protocol is that the agents 
determine whether at least 50\% of them vote ``yes''.

The output function is $O(\qY)=O(\qy)=1$ and $O(\qN)=O(\qn)=0$.
When two agents interact, they change their state according to the following transitions:
\begin{align*}
	a:\ \qY\,\qN &\mapsto \qy\,\qn &
	b:\ \qY\,\qn &\mapsto \qY\,\qy &
	c:\ \qN\,\qy &\mapsto \qN\,\qn &
	d:\ \qy\,\qn &\mapsto \qy\,\qy
\end{align*}
Intuitively, agents are either active ($\qY$, $\qN$) or passive ($\qy$, $\qn$). By transition $a$, when active agents with opposite opinions meet, they become passive. Transitions $b$ and $c$ let active agents change the opinion of passive agents. Transition $d$ handles the case of a tie.

\paragraph{Computations in population protocols.}
Computations use a stochastic model: starting from an initial configuration $C_0$, two agents are repeatedly picked, uniformly at random, and 
the corresponding transition is applied. This gives rise to an infinite sequence $C_0 \trans{t_1} C_1 \trans{t_2} \ldots$ of configurations, 
called a \emph{run}. A run \emph{stabilizes} to consensus $b \in \{0,1\}$ if from some point on all configurations have consensus $b$.
Intuitively, in a run that stabilizes to $b$ the agents eventually agree on the answer $b$.
Given a population protocol $\PP$ and a \emph{predicate} $\varphi$ that maps every configuration $C$ to a value in $\{0,1\}$,  we say that $\PP$  \emph{computes} $\varphi$ if for every initial configuration $C$,
a run starting at $C$ stabilizes to consensus $\varphi(C)$ with probability 1.
The \emph{correctness problem} consists of deciding, given $\PP$ and $\varphi$, whether
$\PP$ computes $\varphi$. Intuitively, a correct protocol almost surely converges to the consensus specified by the predicate. 
Majority Voting is correct and computes the predicate that assigns $1$ to the configurations where initially at 
least 50\% of the agents are in state $\qY$, i.e. we have $\varphi(C) = (C(\qY) \ge C(\qN))$.

\paragraph{Majority Voting is correct.}
To intuitively understand why the protocol is correct, it is useful to split a run into \emph{phases}.  The first phase starts in the initial configuration, and ends when two agents interact using transition $a$ for the last time. Observe that this moment arrives with probability 1 because passive agents can never become active again. Further, at the end of the first phase either all active agents are in state $\qY$, or they are all in state $\qN$. The second phase ends when the agents reach a consensus for the first time, that is, the first time that either all agents are in states $\qY, \qy$, or all are in states $\qN, \qn$. To see that the second phase ends with probability 1, consider three cases.
If initially there is a majority of ``yes'', then at the end of the first phase no agent is in state $\qN$, and at least one is in state $\qY$. This agent eventually moves all passive agents in state $\qn$ to state $\qy$ using transition $b$, reaching a ``yes'' consensus. The case with an initial majority of ``no'' is symmetric. If initially there is a tie, then at the end of the first phase all agents are passive, and transition $d$ eventually moves all agents in state $\qn$ to $\qy$, again resulting in a ``yes'' consensus.
The third phase is the rest of the run. We observe that once the agents reach a consensus no transition is enabled, and so the agents remain in this consensus, proving that the protocol is correct.

\section{Protocol verification with Peregrine 2.0} 
\label{sec:verif}

Peregrine 2.0 allows the user to specify and edit population protocols. (Our running example is
listed in the distribution as \textit{Majority Voting}.) After choosing a protocol, the user can simulate it and gather statistics,
as in Peregrine 1.0~\cite{peregrine}. The main feature of Peregrine 2.0 is its new verification engine based on stage graphs,
which closely matches the ``phase-reasoning'' of the  previous section. 

\begin{figure}[ht]
	\centering
	\begin{minipage}{.43\textwidth}
		\includegraphics[width=1.0\textwidth]{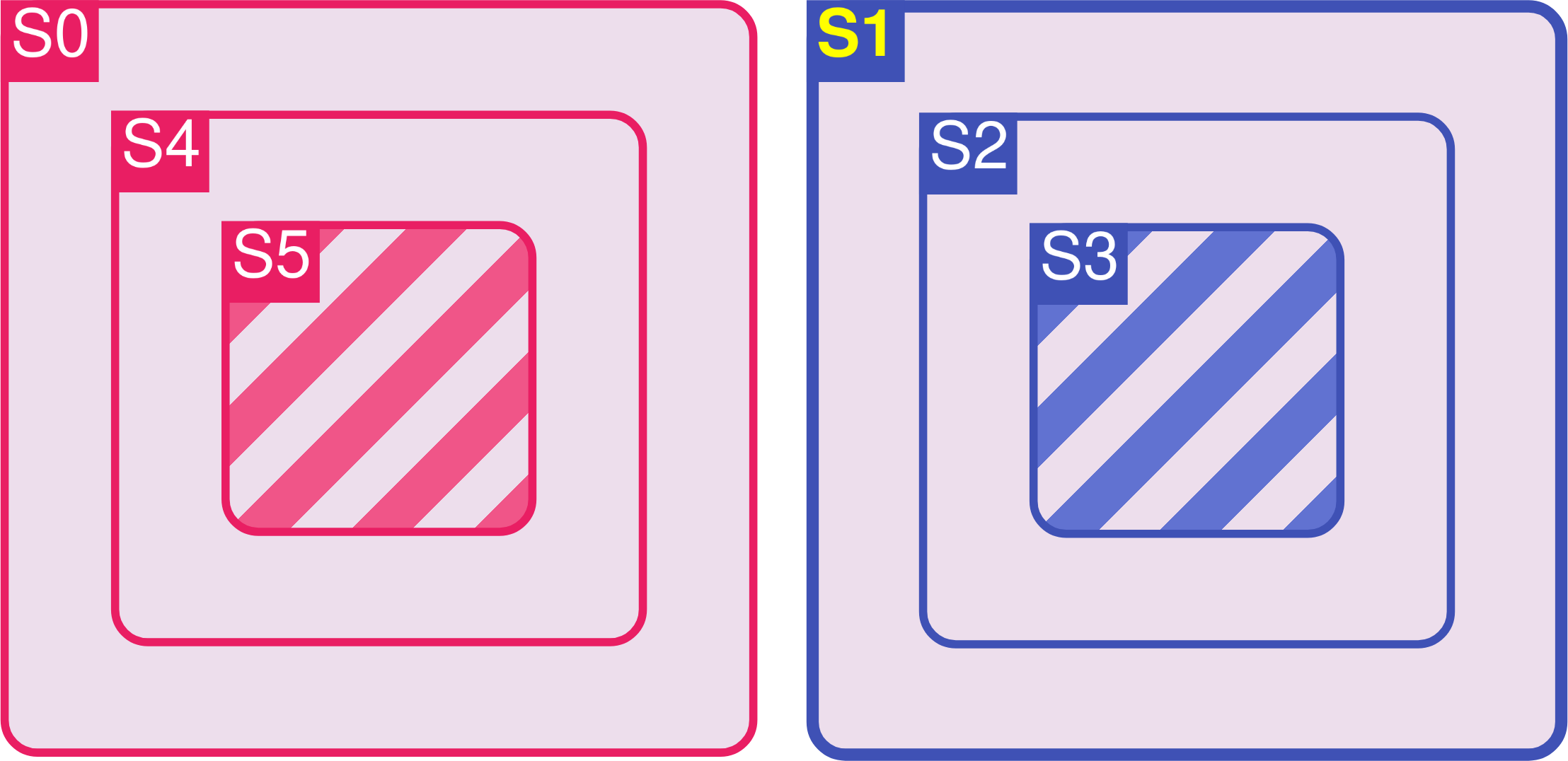}
	\end{minipage}\hfill%
	\begin{minipage}{.56\linewidth}
		\centering
        \begin{tabular}{cl@{\hskip -3pt}c@{\hskip -3pt}c}
            \toprule
            Stage & \multicolumn{1}{c}{Constraint} & Certificate & Speed\\
            \midrule
            $S_0$ & $\mathcal{R}$ & $C(\qY)$ & $\lO(n^2\log n)$\\
            $S_4$ & $\mathcal{R} \land C(\qY) = 0$ & $C(\qy)$ & $2^{\lO(n\log n)}$ \\
            $S_5$ & $\mathcal{R} \land C(\qY){+}C(\qy) = 0$ & $\bot$ & $\bot$ \\
            \midrule
            $S_1$ & $\mathcal{R}'$ & $C(\qN)$ & $\lO(n^2\log n)$ \\
            $S_2$ & $\mathcal{R}' \land C(\qN) = 0$ & $C(\qn)$ & $\lO(n^2\log n)$\\
            $S_3$ & $\mathcal{R}' \land C(\qN){+}C(\qn) = 0$ & $\bot$ & $\bot$ \\
            \bottomrule
        \end{tabular}
	\end{minipage}
    \caption{Stage graphs for Majority Voting protocol with constraints,
    certificates and speeds. The expression
    $\PotReach$ and $\PotReach'$ denote abstractions of the reachability relation,
    which are a bit long and therefore omitted for clarity.}
    \label{fig:stageGraphMajority}
\end{figure}

\paragraph{Stage graphs.} A \emph{stage graph} is a directed acyclic graph whose nodes, called \emph{stages}, are possibly 
infinite sets of configurations, finitely described by a Presburger formula. Stages are \emph{inductive}, 
i.e. closed under reachability. There is an edge $S \rightarrow S'$ to a \emph{child} stage $S'$
if $S' \subset S$, and no other stage $S''$ satisfies $S' \subset S'' \subset S$. Peregrine 2.0 represents stage graphs 
as Venn diagrams like the ones on the left of Figure \ref{fig:stageGraphMajority}.
Stages containing no other stages are called \emph{terminal}, and otherwise \emph{non-terminal}. 
Intuitively, a phase starts when a run enters a stage, and ends when it reaches one of its children. 

Each non-terminal stage $S$ comes equipped with a \emph{certificate}. Intuitively, 
a certificate proves that runs starting at any configuration of $S$ will almost 
surely reach one of its children and,  since $S$ is inductive, get trapped there forever. 
Loosely speaking, certificates take the form of ranking functions bounding the distance of a 
configuration to the children of $S$, and are also finitely represented by Presburger formulas. Given
a configuration $C$ and a certificate $f$, runs starting at $C$ reach a configuration $C'$
satisfying $f(C') < f(C)$ with probability 1. 

To verify that a protocol computes a predicate $\varphi$ we need two stage graphs, one for each output. 
The roots of the first stage graph contain all initial configurations $C$ with $\varphi(C) = 0$
and the terminal stages contain only configurations with consensus 0.
The second handles the case when $\varphi(C) = 1$. 

\begin{figure}[ht]
	\centering
	\begin{minipage}{.49\textwidth}
		\includegraphics[width=1.0\textwidth]{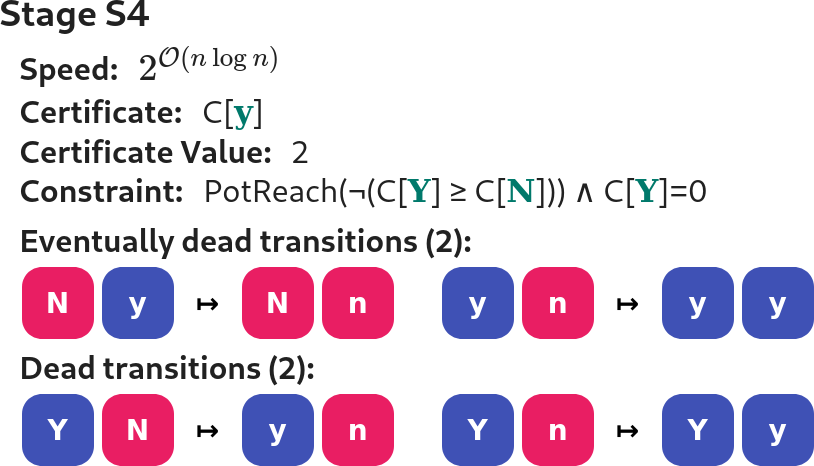}
		\caption{Details of stage $S_4$ in \Cref{fig:stageGraphMajority}
			at configuration $\multiset{\qN,4\cdot \qn,2 \cdot \qy}$.
            The terms $\mathsf{C}[\textcolor{teal700}{\mathtt{q}}]$ are the number of agents $C(q)$ in state $q$.
            }
		\label{fig:stageDetails}
	\end{minipage}\hfill%
	\begin{minipage}{.49\linewidth}
		\centering
		\includegraphics[width=1.0\textwidth]{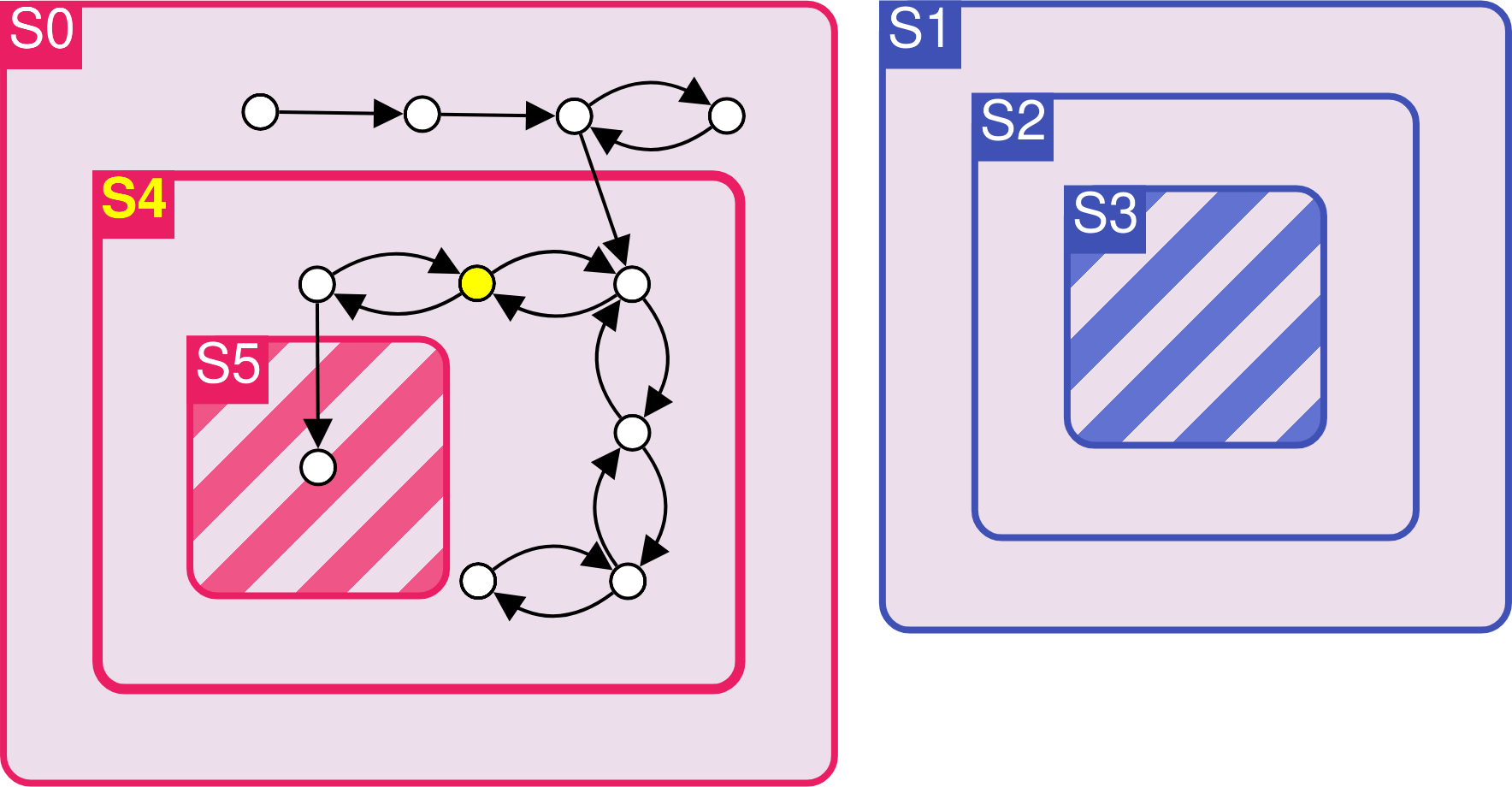}
		\caption{Partially constructed Markov chain after a simulation of the Majority Voting protocol inside the protocol's stage graphs, with \protect\confSelected{} = $\multiset{\qN,4\cdot \qn,2 \cdot \qy}$ selected.}
		\label{fig:visualizingExecutions}
	\end{minipage}
\end{figure}

\paragraph{Stage graphs for Majority Voting.} For the Majority Voting protocol 
Peregrine 2.0 generates the two stage graphs of Figure \ref{fig:stageGraphMajority} in a completely automatic way. 
By clicking on a stage, say $S_4$, the information shown in \Cref{fig:stageDetails} is displayed. The constraint describes the set of configurations of the stage
(Figure \ref{fig:stageGraphMajority} shows the constraints for all stages).
In particular, all the configurations of $S_4$ satisfy $C(\qY)=0$, that is, all agents initially in state $\qY$ have already become passive.
The certificate indicates that a run starting at a configuration $C \in S_4 \setminus S_5$ 
eventually reaches $S_5$ or a configuration $C' \in S_4 \setminus S_5$ such that $C'(\qy) < C(\qy)$. Peregrine 2.0 also
displays a list of \emph{dead transitions} that can never occur again from any configuration of $S_4$, 
and a list of \emph{eventually dead transitions}, which will become dead whenever 
a child stage, in this case $S_5$, is reached. 

While they are automatically generated, these stage graphs closely map the intuition above.
The three stages of each graph naturally correspond to the three phases of
the protocol: $S_0$ and $S_1$ correspond to the first phase (we reduce $C(\qY)$ or $C(\qN)$),
$S_2$ and $S_4$ to the second phase ($C(\qY)$ or $C(\qN)$ is zero, and we reduce $C(\qy)$ or $C(\qn)$), 
and $S_3$ and $S_5$ to the third phase (all agents are in consensus).

\paragraph{Speed.}  Because agents interact randomly, the length of the phase associated to a stage is a random variable
(more precisely, a variable for each number of agents). The expected value of this variable is called 
the \emph{speed} of the stage. A stage has speed $\lO(f(n))$ if for every $n$ the expected
length of the phase for configurations with $n$ agents is at most $c \cdot f(n)$ for some constant $c$. Peregrine 2.0
computes an upper bound for the speed of a stage using the techniques of~\cite{BEK18}. The last column of Figure~\ref{fig:stageGraphMajority} gives the upper bounds
on the speed of all stages. Currently, Peregrine 2.0 can prove one of the bounds $\lO(n^2 \log n)$, $\lO(n^3)$, $\lO(n^k)$ for some $k$ and
$2^{\lO(n \log n)}$. Observe that for stage $S_4$ of Majority Voting the tool returns $2^{\lO(n \log n)}$.
Majority Voting is indeed very inefficient, much faster protocols exist.

\section{Visualizing runs in the stage graph}
\label{sec:visual}
To further understand the protocol, Peregrine 2.0 allows the user to simulate a run and monitor its
progress through the stage graph.
The simulation is started at a chosen initial configuration or a precomputed example configuration of a stage.
The current configuration is explicitly shown and also highlighted as a yellow circle in the stage graph.
To choose the next pair of interacting agents, the user can click on them.
The resulting interaction is visualized, and the successor configuration is automatically
placed in the correct stage, connected to the previous configuration.
After multiple steps, this partially constructs the underlying Markov chain of the system as shown in \Cref{fig:visualizingExecutions}.
One can also navigate the current run by clicking on displayed configurations or using the \button{PREV} and \button{NEXT} buttons.

\begin{figure}[htb]
  \begin{minipage}[c]{0.47\textwidth}
    \vspace{0.31cm}
	\includegraphics[width=\textwidth]{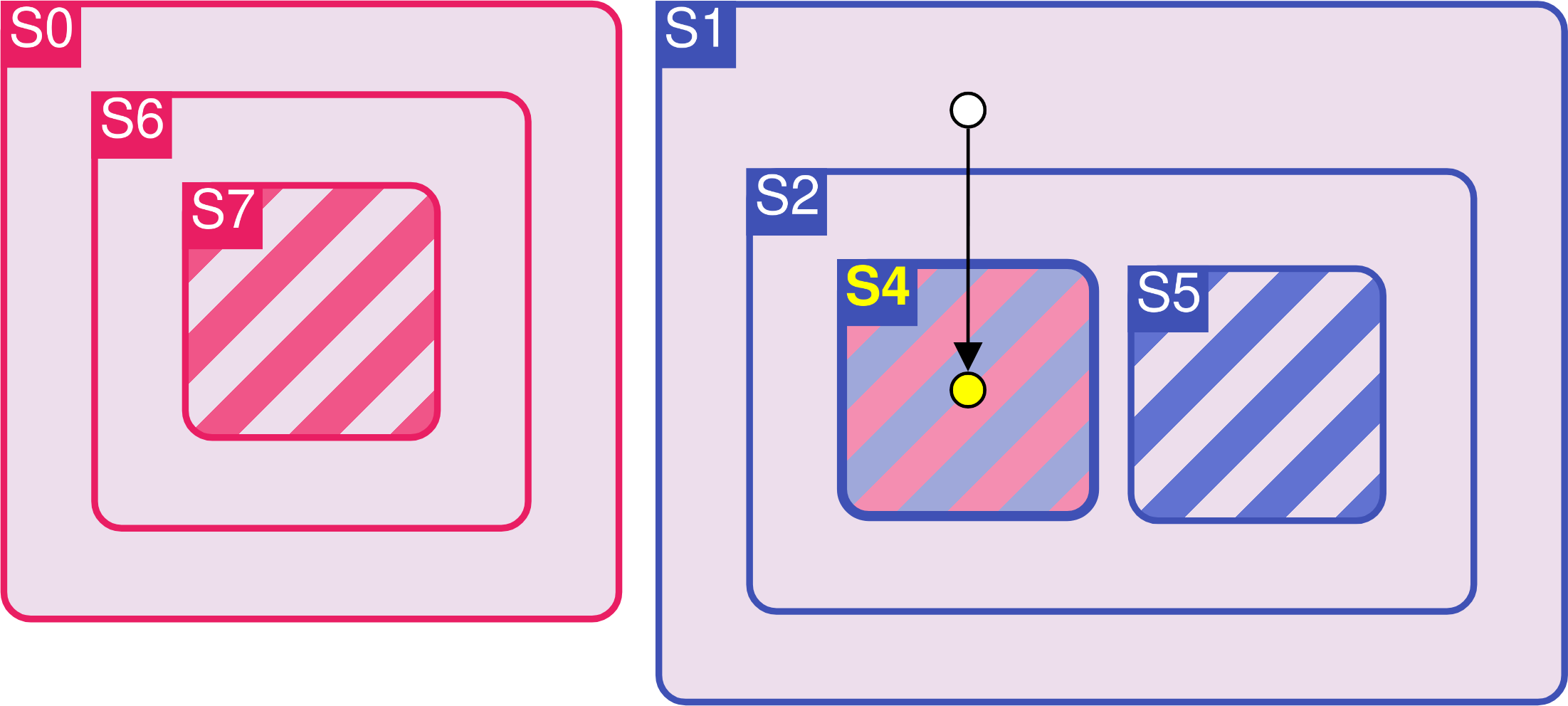}
  \end{minipage}\hfill%
  \begin{minipage}[c]{0.52\textwidth}
      \caption{Counterexample automatically found by Peregrine when verifying Majority Voting (broken), shown in the stage graphs as a run
      from \protect\conf{} = $\multiset{\qY,\qN}$ to \protect\confSelected{} = $\multiset{\qy,\qn}$.
      The graph with root $S_1$ is only a partial stage graph, because stage $S_4$ contains configurations that do not have the correct consensus.}
	\label{fig:counterexample}
  \end{minipage}
\end{figure}

Beyond choosing pairs of agents one by one, the user can simulate a full run of the protocol by clicking on \button{PLAY}.
The acceleration slider allows to speed up this simulation.
However, if the overall speed of the protocol is very slow, a random run might not make progress in a reasonable time frame.
An example for this is the Majority Voting protocol for populations with a small majority for $\qN$, where the expected number of interactions to go from $S_4$ to $S_5$ is $2^{\lO(n \log n)}$.
Thus, even for relatively small configurations like $\multiset{4 \cdot \qY, 5 \cdot \qN}$ a random run is infeasible.
To make progress in these cases, one can click on \button{PROGRESS}.
This automatically chooses a transition that reduces the value of the certificate. Intuitively, reducing the certificate's value guides the run towards a child stage and thus, the run from $S_4$ to $S_5$ needs at most $n$ steps.
To visualize the progress, the value of the stage's certificate for the current configuration is displayed in the stage details as in~\Cref{fig:stageDetails}
and next to the \button{PROGRESS} button.

\paragraph{Finding counterexamples.}
The speed of stage $S_4$ with certificate $C(\qy)$ is so low because of transition $d: \qy\,\qn \mapsto \qy\,\qy$ that increases the value of the certificate
and may be chosen with high probability.
Removing the transition $d$ makes the protocol faster
(this variant is listed in the distribution as ``Majority Voting (broken)'').
However, then Peregrine cannot verify the protocol anymore,
and it even finds a counterexample: a run that does not stabilize to the correct consensus.
\Cref{fig:counterexample} shows the counterexample ending in the configuration $\multiset{\qy,\qn}$ from the initial configuration $\multiset{\qY,\qN}$, i.e. a configuration with a tie.
In this case, the configuration should stabilize to 1, but no transition is applicable at $\multiset{\qy,\qn}$, which does not have consensus 1.
This clearly shows why we need the transition $d$.
Note however that the left part with root stage $S_0$ in \Cref{fig:counterexample} is a valid stage graph, so the modified protocol works correctly in the negative case.
This helps locate the cause of the problem.

\bibliographystyle{splncs04}
\bibliography{bibliography}

\end{document}